\documentclass[12pt,preprint]{aastex}

\shorttitle{Radiative Cooling in Primordial Gas}
\shortauthors{Hirano \& Yoshida}

\newcommand{\HH}{H$_2$ }

\newcommand{\simgt}{\lower.5ex\hbox{$\; \buildrel > \over \sim \;$}}

\def\simge{
    \mathrel{\rlap{\raise 0.511ex
        \hbox{$>$}}{\lower 0.511ex \hbox{$\sim$}}}}
\def\simle{
    \mathrel{\rlap{\raise 0.511ex
        \hbox{$<$}}{\lower 0.511ex \hbox{$\sim$}}}}

\begin{document}

\title{Radiative cooling implementations in simulations of primordial star formation}

\author{Shingo Hirano\altaffilmark{1}}
\email{hirano@astron.s.u-tokyo.ac.jp}
\and
\author{Naoki Yoshida\altaffilmark{2,3}}
\email{naoki.yoshida@ipmu.jp}

\altaffiltext{1}{Department of Astronomy, School of Science, 
University of Tokyo, Bunkyo, Tokyo 113-0033, Japan}
\altaffiltext{2}{Department of Physics, School of Science, 
University of Tokyo, Bunkyo, Tokyo 113-0033, Japan}
\altaffiltext{3}{Kavli Institute for the Physics and Mathematics of the Universe, 
TODIAS, University of Tokyo, Kashiwa, Chiba 277-8583, Japan}

\begin{abstract}
We study the thermal evolution of primordial star-forming gas clouds 
using three-dimensional cosmological simulations.
We critically examine 
how assumptions and approximations made in 
calculating radiative cooling rates 
affect the dynamics of the collapsing gas clouds.
We consider two important molecular hydrogen cooling processes 
that operate in a dense primordial gas; 
${\rm H_2}$ line cooling and 
continuum cooling by ${\rm H_2}$ collision-induced emission.
To calculate the optically thick cooling rates, 
we follow the Sobolev method for the former, 
whereas we perform ray-tracing for the latter.  
We also run the same set of simulations 
using simplified fitting functions for the net cooling rates.
We compare the simulation results in detail. 
We show that the time- and direction-dependence of hydrodynamic quantities
such as gas temperature and local velocity gradients 
significantly affects the optically thick cooling rates.
Gravitational collapse of the cloud core is accelerated 
when the cooling rates are calculated by using the fitting functions. 
The structure and evolution of the central pre-stellar disk 
are also affected.
We conclude that 
physically motivated implementations of radiative transfer 
are necessary to follow accurately 
the thermal and chemical evolution of 
a primordial gas to high densities. 
\end{abstract}


\section{Introduction}

The first stars fundamentally transform the early universe 
by emitting the first light and also 
by synthesizing and dispersing the first heavy elements. 
They initiate cosmic reionization, and set the scene for 
the subsequent formation of the first galaxies 
(see \cite{bromm11} for a review).
Understanding how and when the first stars were formed is 
one of the important goals of modern astronomy.

There has been a significant progress over the past years 
in the theoretical studies on the first stars.
With the currently available computer power,
one can perform an {\it ab initio} simulation of the first star formation 
in substantial details (see \cite{bromm09} for a review).
\cite{yoshida08} studied the formation of a primordial protostar 
in a proper cosmological context.
\cite{turk09} showed that 
a primordial gas cloud can fragment into multiple clumps 
by the action of the rotation and turbulence in the cloud. 
\cite{clark11} and \cite{greif11} 
used sink-particle techniques to follow 
the evolution of the accretion disk around a primordial protostar 
for hundreds years.
It was shown that 
the circumstellar disk around a protostar becomes 
gravitationally unstable to form multiple protostars. 
Unfortunately, these simulations could not be run long enough to obtain 
the solid prediction for multiplicity and 
for the characteristic mass of the first stars.
Including the so-called proto-stellar feedback effects 
is important to determine the mass of the first stars \citep{mckee08, stacy12}.
Recently, \cite{hosokawa11} performed 
radiation-hydrodynamical calculations to show that 
the self-regulating proto-stellar feedback halts 
the growth of a primordial protostar 
when its mass is several tens of solar-masses. 
It has become possible to follow the entire evolution
of a primordial protostar to the main-sequence and thus to
discuss rigorously important issues such as the characteristic
mass of the first stars.

There still remain a few uncertainties 
in these theoretical studies 
on primordial star formation.
For example, 
some of the important chemical reaction rates in a pre-stellar gas 
are not known to good accuracies \citep{glover08}.
Most importantly, 
the three-body hydrogen molecule formation rate 
is poorly determined, which leaves substantial uncertainties
in the thermal evolution of a pre-stellar gas 
at high densities \citep{turk11}.
Calculating radiative cooling rates at such high densities, 
where the gas is optically thick, 
is essentially a radiative transfer problem. 
Previous studies adopt two methods to solve this problem.
Fitting functions are proposed by \cite{ripamonti04}
which describes the net cooling rate 
as a function of the local density based on the result of 
a fully one-dimensional (1D) radiative transfer calculation.
The other method adopts the large-velocity gradient approach, 
the so-called Sobolev method \citep{yoshida06, clark11, hosokawa11}.
Similar methods have been used to follow the gas evolution 
at even higher densities \citep{ripamonti04, yoshida07, yoshida08}.
It is important to examine whether or not 
the different implementations of radiative cooling calculations
produce significantly different results.

In this paper, we critically examine 
whether or not implementations of the optically-thick radiative cooling 
affect the evolution of a primordial gas cloud.
To this end, we run a set of 
three-dimensional (3D) cosmological hydrodynamical simulations. 
We explicitly compare the results obtained from the simulations and 
study the structure of the gas cloud in detail.

The rest of the paper is organized as follows.
In Section 2, we describe 
the main physical processes in a primordial gas cloud evolution.
There, we also describe the calculation methods of the net cooling rates. 
The simulation settings and computational methods are given in Section 3. 
Section 4 shows results of the cosmological simulations. 
We summarize the results and give concluding remarks in Section 5.

\section{Thermal evolution of a primordial gas cloud}

A gravitationally contracting primordial gas 
evolves roughly isothermally.
Since the onset of run-away collapse, 
the temperature rises only a factor of about ten 
whereas the density increases 
over 16 orders of magnitudes \citep{palla83, omukai98}.
In a primordial gas, 
the main coolant is hydrogen molecules ($\rm {H_2}$). 
There are two important regimes 
where radiative transfer effects become important.
One is at densities 
$10^8 \ {\rm cm^{-3}} \simle n_{\rm H} \simle 10^{14} \ {\rm cm^{-3}}$,
where ${\rm H_2}$ line cooling is dominant and 
the cloud core cools and condenses rapidly. Then the cloud core
becomes optically-thick to ${\rm H_2}$ lines. 
It is known that the chemo-thermal instability can be triggered 
in this phase \citep{sabano77, silk83}.
The other is at densities 
$10^{14} \ {\rm cm^{-3}} \simle n_{\rm H} \simle 10^{17} \ {\rm cm^{-3}}$,
where the gas is nearly opaque to ${\rm H}_2$ line photons
but cooling by the collision-induced emission (CIE) becomes efficient.

In order to compute the net radiative cooling rate in the two regimes,
we need to compute the opacity for photons in a broad energy range.
In principle, the gas opacity depends on 
a number of physical quantities 
such as the local gas density, temperature, and velocities.

\subsection{${\rm H_2}$ line cooling}

When the gas density exceeds ${\rm n_H} \sim 10^{8} \ {\rm cm}^{-3}$,
rapid three-body reactions of \HH formation convert 
nearly all the hydrogen atoms into hydrogen molecules.
Then \HH line cooling becomes highly efficient. 
As the gas cloud condenses, however, 
the gas cloud core becomes opaque to \HH lines.
Then the \HH line cooling rate is calculated as
\begin{equation}
\Lambda_{\mathrm{H_2, thick}} = \sum_{u, l} h \nu_{ul} \ \beta_{{\rm escape}, ul} \ A_{ul} \ n_u \ ,
\label{eq:H2_line}
\end{equation}
where 
$h \nu_{ul}$ is the energy difference between the upper level $u$ and the lower level $l$,
$\beta_{\mathrm{escape}, ul}$ is the escape probability for a photon without absorption, 
$A_{ul}$ is the Einstein coefficient for the spontaneous transition, and 
$n_u$ is the number density of the hydrogen molecule in the upper level $u$.
To calculate the escape probability $\beta_{\rm escape}$, 
we evaluate the opacity for each \HH line as
\begin{equation}
\tau_{lu} = \alpha_{lu} L \ ,
\label{eq:line_opacity}
\end{equation}
where 
$\alpha_{lu}$ is the absorption coefficient 
for the transition from $l$ to $u$ level.
Calculating the characteristic absorption 
length scale, $L$, is the remaining task.
To this end, we adopt the Sobolev method. 
The Sobolev length along a line of sight is defined as
\begin{equation}
L_r = {{v_{{\rm thermal}}}\over{|dV_r/dr|}} \ ,
\label{eq:Sobolev_length}
\end{equation}
where 
$v_{\mathrm{thermal}} = (kT/m_\mathrm{H})^{1/2}$ is 
the typical thermal velocity of the hydrogen molecules and 
$V_r$ is the fluid velocity in the direction.
The escape probability for a spherical cloud is given by
\begin{equation}
\beta_{\mathrm{escape}} = {{1 - \exp(-\tau)}\over{\tau}} \ , \ \tau = \alpha L_r \ .
\label{eq:escape_probability}
\end{equation}
We compute the Sobolev length and 
the escape probability in three orthogonal directions 
and use the average value of them to calculate the net cooling rate.

The above method requires 
the evaluation of the optical depths 
for a few hundred molecular lines, 
and thus is computationally costly.
\cite{ripamonti04} propose a fitting formula 
for the optically-thick cooling rate based on
the result of 1D radiative transfer simulations.
It is given by
\begin{equation}
\Lambda_{{\rm H}_2, {\rm thick}}=\Lambda_{{\rm H}_2, {\rm thin}} \times 
\min \left[ 1, \left(\frac{n}{8 \times 10^9 \ {\rm cm}^{-3}}\right)^{-0.45} \right] \ .
\label{eq:line_cooling_rate_thick}
\end{equation}
Note that the simple function 
depends only on the local gas density.
We use the formula as an alternative method 
to compute the net cooling rate.

\subsection{Collision-induced emission cooling}

At densities greater than 
$n_{\rm H} \sim 10^{14} \ {\rm cm^{-3}}$, 
hydrogen molecules collide so frequently that 
each collision pair temporarily induces an electric dipole and 
either molecule makes an energy transition by emitting a photon.
This process is known as collision-induced emission (CIE),
which acts as an efficient radiative cooling process 
at such high densities. The resulting emission spectrum appears 
essentially as continuum radiation. 
We use the collision cross-sections of 
\cite{jorgensen00}, \cite{borysow01}, and \cite{borysow02}.

At even higher densities 
$n_{\rm H} > 10^{16} \ {\rm cm^{-3}}$, 
the gas cloud becomes opaque to the continuum emission. 
We use the Planck opacity table \citep{lenzuni91} 
to calculate the gas opacity and 
the resulting radiative cooling rate.
By noting that the net energy transfer rate should scale as 
$\Lambda \propto 1/(1+\tau)$ for small $\tau$, 
whereas it scales as 
$\Lambda \propto 1/{\tau}^2$ for large $\tau$,
we assume a simple form of double power-law
\begin{equation}
f = \frac{1}{[1+\tau(n, T)][1+(\tau(n, T)/10)]}  
\label{eq:continuum_opacity}
\end{equation}
as the ``efficiency'' factor of the CIE cooling. 
Although the particular functional form is somewhat {\it ad hoc},
it reproduces well the net cooling rate 
that is obtained from more detailed radiative transfer calculations
\citep{omukai98, yoshida08}. 
Note that 
we explicitly write as $\tau = \tau(n, T)$ to express that
the local opacity depends on the gas density and temperature.
In practice, we compute the opacity along a line-of-sight by the integral
\begin{equation}
\tau = \int \kappa(\rho, T) \rho \;{\rm d}l \ ,
\end{equation}
where 
$\kappa$ is the absorption coefficient and
$\rho$ is the local gas density. 
In order to take the cloud geometry and structure into account, 
we take the mean of the efficiency factors in three orthogonal directions 
(see equation [\ref{eq:continuum_opacity}]),
\begin{equation}
f_{{\rm mean}} = \frac{f_{\rm x} + f_{\rm y} + f_{\rm z}}{3} \ .
\label{eq:fmean}
\end{equation}
Then the CIE cooling rate is calculated as 
\begin{equation}
\Lambda_{\mathrm{CIE, thick}} = \Lambda_{\mathrm{CIE, thin}} \times f_{{\rm mean}} \ .
\label{eq:H2_CIE}
\end{equation}

A simple continuum opacity model 
is also proposed by \cite{ripamonti04},
which is given by a function of 
the local gas density $n_{{\rm H}_2}$ as
\begin{equation}
\Lambda_{{\rm CIE}, {\rm thick}}=\Lambda_{{\rm CIE}, {\rm thin}} \times 
\min \left[ 1, \frac{1 - e^{- {\tau}_{\rm CIE}}}{{\tau}_{\rm CIE}} \right] \ ,
\label{eq:cie_cooling_rate_thick}
\end{equation}
where
\begin{equation}
{\tau}_{\rm CIE} = \left( \frac{n_{{\rm H}_2}}{7 \times 10^{15} \ {\rm cm^{-3}}} \right)^{2.8} \ .
\label{eq:cie_tau}
\end{equation}

We compare the runs with the above two methods 
of computing the CIE cooling rate. 
(as equation [\ref{eq:H2_CIE}] and [\ref{eq:cie_cooling_rate_thick}]).

\section{Numerical Simulations}

We use the parallel $N$-body/Smoothed Particle Hydrodynamics (SPH) solver 
$GADGET$-2 \citep{springel05} in its version 
suitably adopted for the primordial star formation.
We solve chemical rate equations 
for fourteen species of primordial species
(${\rm e^-}$, ${\rm H}$, ${\rm H^+}$, ${\rm He}$, ${\rm He^+}$, 
${\rm He^{++}}$, ${\rm H_2}$, ${\rm H_2^+}$, ${\rm H^-}$, ${\rm D}$, 
${\rm D^+}$, ${\rm HD}$, ${\rm HD^+}$, ${\rm HD^-}$).
The reactions and rates are summarized in \cite{yoshida06, yoshida07}.
We adopt the $\Lambda$-Cold Dark Matter ($\Lambda$-CDM) cosmology.
The cosmological parameters are based on 
the seven-year WMAP results \citep{larson11}; 
${\Omega}_{\Lambda}=0.734, \ 
{\Omega}_{m}=0.236, \ 
{\Omega}_{b}=0.0449$, and 
$H_{0}=71.0 \ \mathrm{km \ s^{-1} \ Mpc^{-1}}$.
The normalization of the power spectrum 
is set to be ${\sigma}_8 = 2.0$ 
such that structure forms early in the small simulation volume.
All simulations are initialized at $z_{\rm ini} = 99$.

We use the zoom-in re-simulation techniques 
to achieve a large dynamic range. 
First, we run a cosmological simulation 
to locate a star-forming halo 
which is later re-simulated with a higher resolution.
The comoving box size of this parent simulation is
$50 \ {\rm kpc \ h^{-1}}$ on a side. 
In the zoomed region, 
the initial particle masses are 
$m_{\rm gas} = 2.04 \times 10^{-3} \ M_{\odot}$ and 
$m_{\rm DM} = 1.01 \times 10^{-2} \ M_{\odot}$, respectively.

We run each of the zoomed simulations first to the point where 
the central density reaches $n_{\rm cen} = 10^8 \ {\rm cm^{-3}}$.
Then, to calculate the gas collapse further, 
we use the particle split method of \cite{kitsionas02} 
to achieve a higher mass resolution.
We do this refinement progressively such that 
a local Jeans length is always resolved 
by 10 times the local SPH smoothing length. 
By using this technique, 
we achieve a mass resolution of 
$m_{\rm gas} = 4 \times 10^{-6} \ M_{\odot}$ 
at the last output time. 
We stop the simulations 
when the central density reaches 
$n_{\rm cen} = 10^{17} \ {\rm cm^{-3}}$. 

\section{Results}

\begin{figure}
\begin{center}
\resizebox{16cm}{!}{\includegraphics{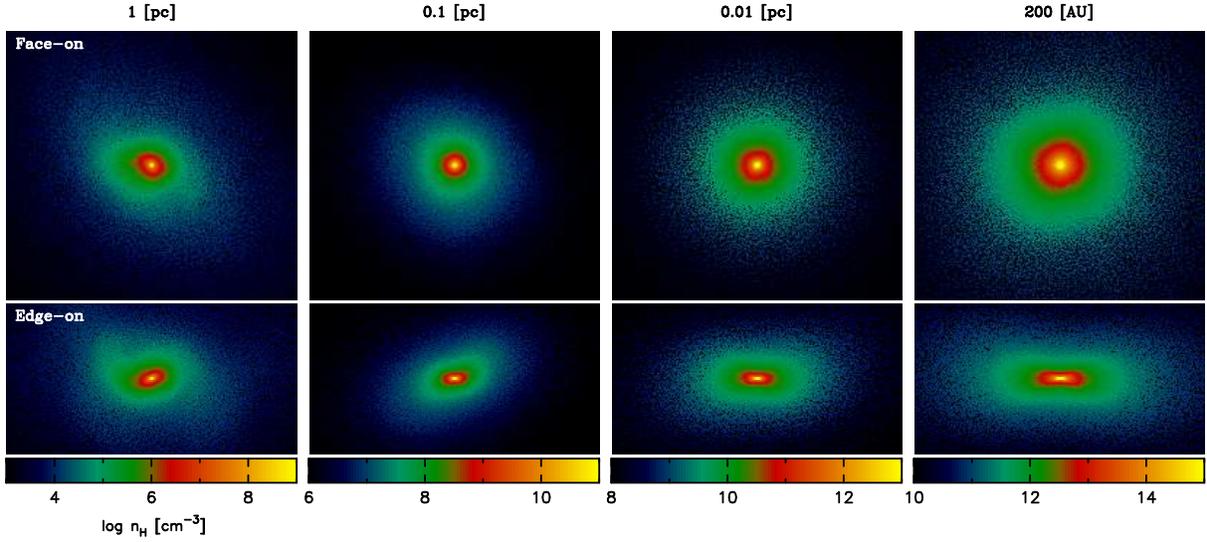}}
\caption{
Projected density distribution for Run A, 
which shows a flattened disk-like structure.
Face-on views (top panels) and 
edge-on views (bottom panels) in a volume of, 
from left to right, 
$1$, $10^{-1}$, $10^{-2}$, and 
$10^{-3}$ pc $\sim 200$ AU on a side, respectively.}
\label{f1}
\end{center}
\end{figure}

To examine the properties of star-forming halos, 
we choose two realizations as characteristic cases, Run A and B.
Run A forms a disk-like structure 
inside the collapsing gas cloud, 
whereas Run B shows elongated structure and 
eventually develops S-shape arms
(see the top panels in Figures \ref{f1} and \ref{f2}). 
The bottom panels show, in both cases, 
the central sub-parsec region is significantly flattened. 
The non-spherical structures likely yield 
direction-dependence of both line and continuum opacities.

In the following subsections, 
we first compare the opacity models for \HH line cooling.
Then we examine differences in the CIE cooling phase.
We have already mentioned the importance of 
the three-body molecular hydrogen formation rate \citep{turk11}. 
We examine the effect of varying the three-body reaction rate
at the end of this section.

\subsection{${\rm H_2}$ line cooling}

\begin{figure}
\begin{center}
\resizebox{12cm}{!}{\includegraphics{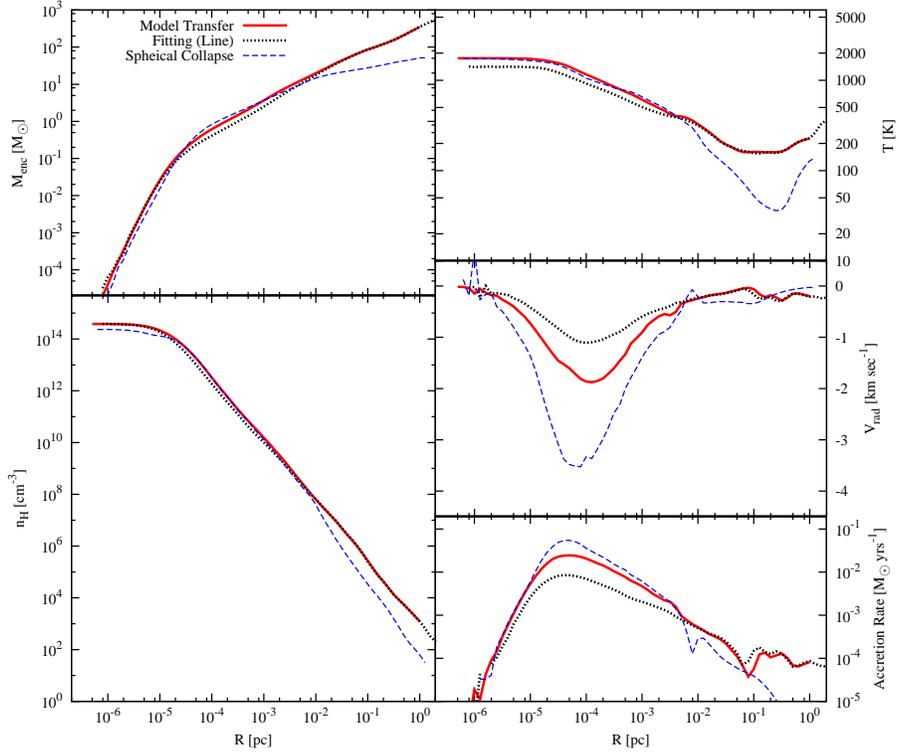}}
\caption{
Radial profiles of various quantities for Run A. 
Five panels show 
the enclosed mass, temperature, radial velocity,
gas number density, and gas mass accretion rate, 
in the clockwise order starting from the top-left.
The solid line shows the result with 
the 3D ${\rm H_2}$ line 
opacity calculation (Sobolev method), 
the dotted line is for the run with the fitting function 
and the dashed line is for 
a spherical collapse calculation with 3D set-up.
}
\label{f3}
\end{center}
\end{figure}

Figure \ref{f3} shows
the radial profiles of five physical quantities for Run A
when the central density is 
$n_{\rm H} \sim 10^{14} \ {\rm cm^{-3}}$.
At this time, the central part is nearly completely opaque 
to \HH line emission. 
The solid line shows the result from the run with 
the 3D opacity calculation 
(Sobolev method, equation [\ref{eq:H2_line}]) and 
the dotted line is for the run with the fitting opacity formula 
(equation [\ref{eq:line_cooling_rate_thick}]).
We see significant differences in  
the temperature, radial velocity, and accretion rate 
profiles.

Because the photon escape probability 
depends on the local velocity gradient 
(see equation [\ref{eq:Sobolev_length}]), 
the line cooling rate can be large 
if the velocity gradient is large in the dense cloud core.
The radial infall velocity is critically affected by
the degree of rotation of the collapsing cloud.
Star-forming clouds in a cosmological simulation generically
have finite initial angular momenta and thus they spin up
gradually as they collapse gravitationally. 
The radial infall velocity, $V_r$, and the velocity gradient, $dV_r/dr$, 
in such rotating gas clouds are smaller 
than realized in spherically symmetric collapse. 
Thus the escape probability in the cosmological simulation 
is smaller than the fitting formula predicts.

In order to investigate further the difference in the escape probability, 
we perform a spherical collapse simulation with 3D set-up. 
We follow gravitational collapse of a super-critical Bonnor-Ebert sphere 
having a mass of $\sim 1000 \ M_{\odot}$. 
For this run, we calculate the \HH line opacity using the Sobolev method. 
The results are shown in Figure \ref{f3} 
(dashed line in each panel). 
Clearly, the radial velocity of the run
is larger than that of our cosmological simulation Run A, 
which has a substantial degree of rotation. 
This is the major source of the differences in the line 
escape probability and 
in the cooling rate, as shown in Figure \ref{f4}.

The left panel of Figure \ref{f4} shows 
the escape probability of \HH line photons 
as a function of the gas density.
We show three snapshots for the mean profiles (solid lines)
when the central density is
$n_{\rm cen} = 10^{10}, \ 10^{12}$, and $10^{14} \, {\rm cm}^{-3}$.
The escape probability calculated
by our 3D treatment is
smaller than the fitting formula (dotted line) most of the time.
The difference is as large as a factor of ten at the densest part. 
The fitting formula 
over-estimates the net cooling rate. This is easily understood
by the collapse speed of the spherically symmetric calculation. 

Next, let us consider 
the direction-dependence of the escape probability.
In the left panel of Figure \ref{f4}, 
we plot the escape probability 
in the direction along $x-,\ y-,$ and $z$-axes 
(long-dashed, short-dashed, and dot-dashed lines). 
We configure the coordinate such that the $z$-direction is 
aligned to the angular momentum vector of the central cloud core
\footnote{Figures \ref{f1} and \ref{f2} also use the same coordinate.}. 
The escape probability is large along the $z$-direction. 
Thus, physically, line photons preferentially escape 
in perpendicular directions to the flattened cloud core.

It is important to note that the difference (or mis-estimate) 
in the cooling rate critically affects the collapse dynamics.
The right panel of Figure \ref{f4} shows 
the time evolution of the central density 
since the central density reaches 
$n_{\rm cen} = 10^{8} \ {\rm cm}^{-3}$. 
In the run with the fitting formula,
the cloud core collapses earlier by $\sim 20,000$ years,
i.e., the gravitational collapse is accelerated 
because of the ``efficient'' cooling.

\begin{figure}
\begin{center}
\resizebox{12cm}{!}{\includegraphics{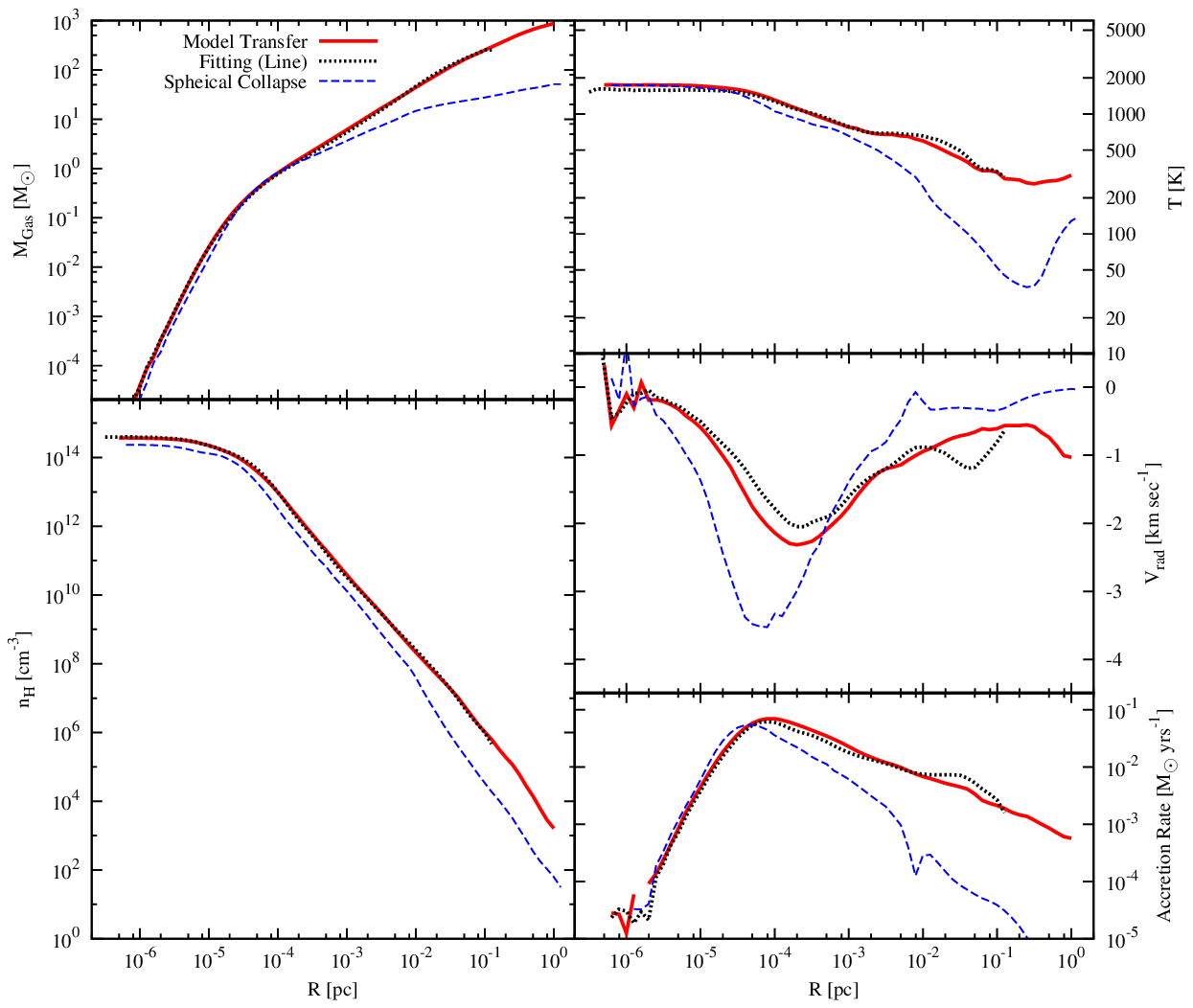}}
\caption{
As for Figure \ref{f3} but for Run B.
}

\label{f5}
\end{center}
\end{figure}

Figures \ref{f5} and \ref{f6} show the same results 
for Run B, which has a spiral structure. 
The overall evolutionary trend is quite similar to Run A. 
Clearly, the importance of radiative transfer effects 
is not particular to the configuration of Run A. 
The differences in the radial profiles of Run B
are understood similarly to Run A as explained in the present
section. 
We conclude that 
the multi-dimensional treatment for the radiative cooling 
is important to follow 
the thermal evolution and the gravitational collapse accurately.

\subsection{${\rm H_2}$ collision-induced emission cooling}

\begin{figure}
\begin{center}
\resizebox{12cm}{!}{\includegraphics{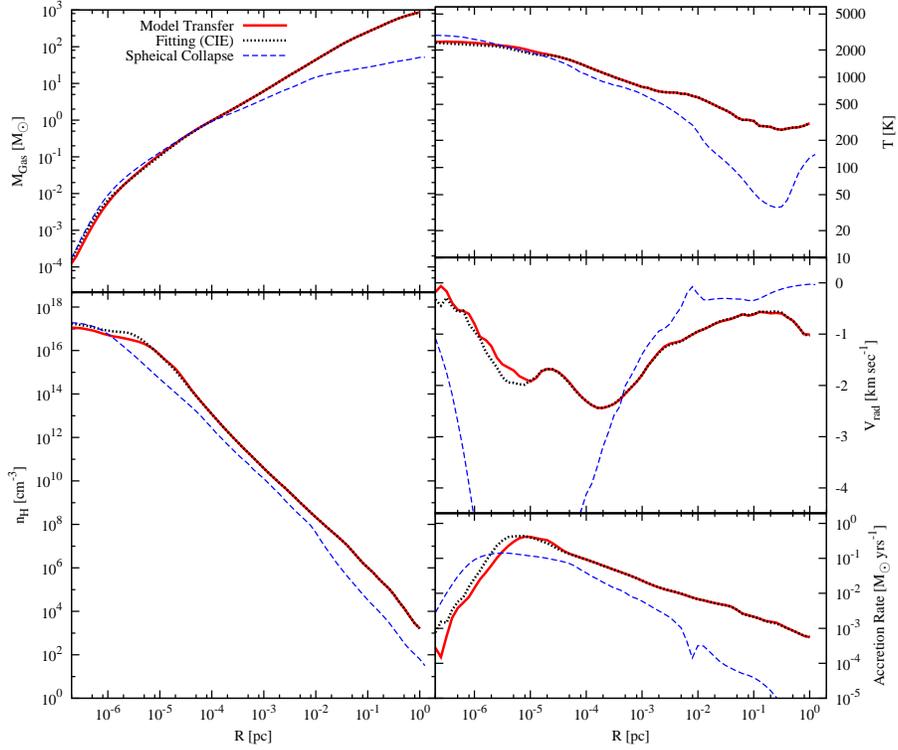}}
\caption{
Radial profiles of various quantities for Run B
in the high density regime where CIE cooling is important.
Five panels show the enclosed mass, temperature, radial velocity, 
gas number density, and mass accretion rate,  
in the clockwise order starting from the top-left.
The solid line is for the result with 
the 3D ${\rm H_2}$ CIE opacity calculation, 
the dotted line is for the run with the fitting function 
and the dashed line is for a spherical collapse calculation 
with 3D treatments.
}
\label{f7}
\end{center}
\end{figure}

Next, we discuss the thermal evolution through the phase 
where CIE cooling is important. 
In this subsection, we primarily discuss the results of Run B. 
The radial profiles when the central density is 
$n_{\rm cen} \sim 10^{17} \ {\rm cm}^{-3}$
are shown in Figure \ref{f7}. 
We also plot the normalized cooling rate, the efficiency factor
$f = \Lambda_{\mathrm{CIE, thick}} / \Lambda_{\mathrm{CIE, thin}}$ 
and the time evolution of the central density in Figure \ref{f8}.

The left panel of Figure \ref{f8} shows 
similar features to the case of ${\rm H_2}$ line opacity discussed
in the previous section.
We a significant direction-dependence of the
escape probability.
The fitting formula over-estimates the net cooling rate 
compared with the run with our 3D opacity treatment.
The difference in the net cooling rate 
becomes as large as a factor of 5 at 
$n_{\rm cen} \sim 5 \times 10^{15} \ {\rm cm}^{-3}$. 
The gas collapses faster
with the fitting formula for the opacity calculation.
However, the difference in the resulting collapse time 
is not very large (see the right panel of Figure \ref{f8}).
The time difference is only about 1 year 
at the end of the calculations.

\begin{figure}
\begin{center}
\resizebox{12cm}{!}{\includegraphics{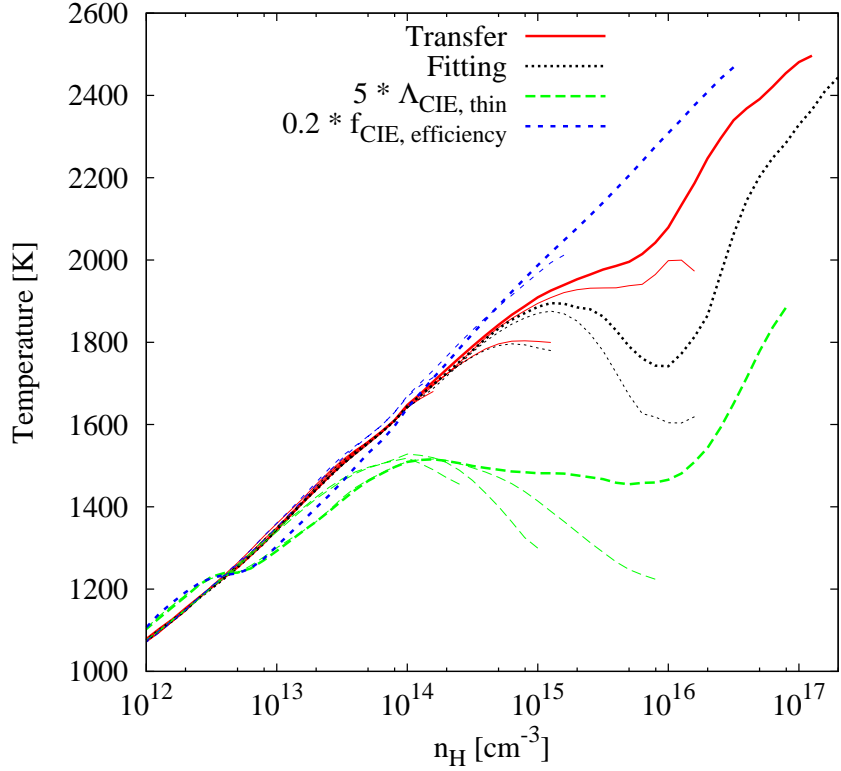}}
\caption{
Thermal evolution in the temperature - density plane for Run B. 
We plot the simulation results 
with our 3D radiative transfer treatment (solid),  
with the fitting opacity function (dotted), 
with a increased CIE cooling rate (long-dashed), and 
with a reduced CIE escape probability (short-dashed).
The thick lines show the final output time and 
the the thin lines are for the results at earlier phases.
}
\label{f9}
\end{center}
\end{figure}

It is worth pursuing the reason why the collapse proceeds 
similarly in the two cases 
despite the large difference in the net cooling rate 
in the relevant regime 
$n_{\rm H} \sim 10^{15} - 10^{16} \ {\rm cm}^{-3}$. 
To this end, we perform two additional calculations for Run B. 
One is run with an artificially increased CIE cooling rate as
$\Lambda_{\rm CIE} \to 5 \times \Lambda_{\rm CIE}$ 
whereas the other is run with the reduced efficiency factor
$f_{\rm mean} \to 0.2 \times f_{\rm mean}$. 
The other configurations are identical to Run B, and
the opacity calculations are also done using 3D ray-tracing.
Therefore, we expect that the two runs clarify the impact
of enhanced/reduced CIE cooling. 
The results are shown in Figure \ref{f9}. 
In the case with the increased cooling rate, 
the collapsing gas evolves on a lower temperature track 
(the long-dashed line in Figure \ref{f9}), as 
is naively expected from the enhanced cooling rate.
Clearly, the cooling rate itself directly affects 
the gas temperature in this regime.
On the other hand, the effect of decreasing the efficiency factor 
appears relatively small (the short-dashed line).
The cloud evolves on a slightly higher temperature track, 
but the difference is significant
only in a narrow range of density, 
$ 10^{15} < n_{\rm H} < 10^{16} \ {\rm cm}^{-3}$,
where the gas is collapsing rapidly. 
The free-fall time there is estimated to be $t_{\rm ff} \sim 1 \ {\rm year}$
which is comparative to the cooling time. 
The core condenses quickly and becomes optically thick 
to the continuum photons after the central density reaches 
$\sim 10^{16} \ {\rm cm}^{-3}$.
In the left panel of Figure \ref{f8}, 
we also plot the time evolution of the cooling efficiency 
at the central part (double-dotted line). 
The evolution looks similar to the fitting function 
at the high density.

In summary, the accuracy of the continuum opacity calculation 
causes only minor effect on the thermal evolution. 
The difference in $f = \Lambda_{\mathrm{CIE, thick}} / \Lambda_{\mathrm{CIE, thin}}$ 
is large between the methods, but the resulting evolution 
is not sensitive to the details of the methods.
It should be noted, however, that we have considered only a particular case
in which the central core undergoes rapid run-away collapse. 
In other circumstances, for example in an accretion disk around 
a protostar where the density evolution is much slower than in
the collapsing gas core that we have studied, 
accurate calculations of the optically-thick cooling may be more important.

\subsection{Three-body ${\rm H_2}$ formation}
 
It is well known that 
there is a large uncertainty 
in the reaction rate of the three-body \HH formation 
\begin{equation}
{\rm H + H + H \to H_2 + H} \ .
\label{eq:three-body}
\end{equation}
Because this is the dominant reaction to form hydrogen molecules 
at high densities ($n_{\rm H} > 10^8 \ {\rm cm^{-3}}$),
an accurate reaction rate is needed to determine
the chemical and thermal evolution of a primordial gas cloud.
\cite{glover08} summarize various rate coefficients used in the literature, 
which differ by a factor of 30 
at the relevant temperature range. 
\cite{turk11} perform a set of hydrodynamical simulations
to directly study the overall effect caused by the uncertainties 
in the reaction rate.
They conclude that, 
while the difference between different realizations (gas cloud samples) 
is larger than that caused by the uncertainty of the three-body 
rate coefficient, the morphology and the collapse time 
of a gas cloud depend strongly
on the reaction rate. 

\begin{figure}
\begin{center}
\resizebox{9.2cm}{!}{\includegraphics{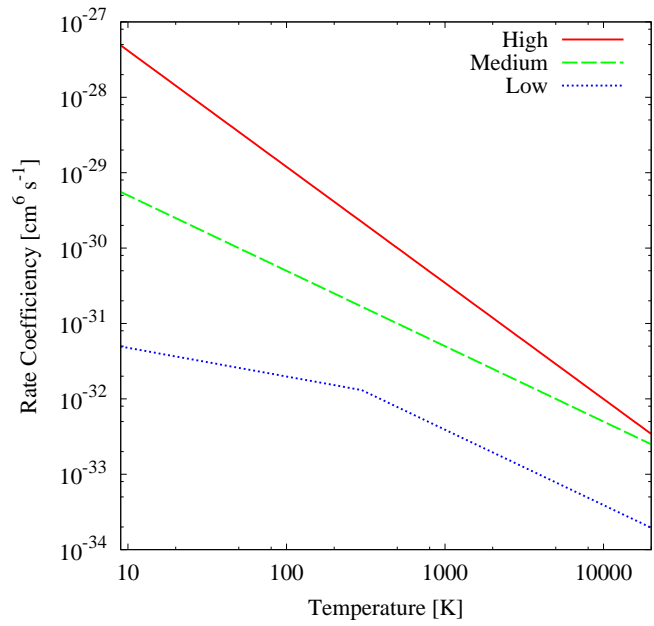}}
\caption{
Three-body \HH formation reaction rate 
used in the literature. 
The solid line (High) is the rate from \cite{flower07}, 
the dashed line (Medium) is from \cite{palla83}, 
and the dotted line is from \cite{abel02}. 
The rates are explicitly given in Table \ref{t1}.
}
\label{f10}
\end{center}
\end{figure}

In this section, we revisit the issue
because the density range where the uncertainties are relevant
is coincident with the range where
the radiative transfer treatment is important, as studied in 
the previous sections.
We are able to compare the overall differences 
caused by the uncertainties of the reaction rates
with the differences caused by radiative transfer treatments.
So far in the present paper,
we have adopted the reaction rate from \cite{palla83} 
($Medium$ case in Figure \ref{f10}).
We run the same simulation as Run B but with
the different three-body reaction rates 
of \cite{flower07} ($High$) and \cite{abel02} ($Low$), 
which are the largest and smallest reaction rates 
among those compiled by \cite{glover08}.
The respective rates are summarized in Figure \ref{f10} and Table \ref{t1}. 

\begin{figure}
\begin{center}
\begin{tabular}{cc}
\includegraphics[clip,scale=0.6]{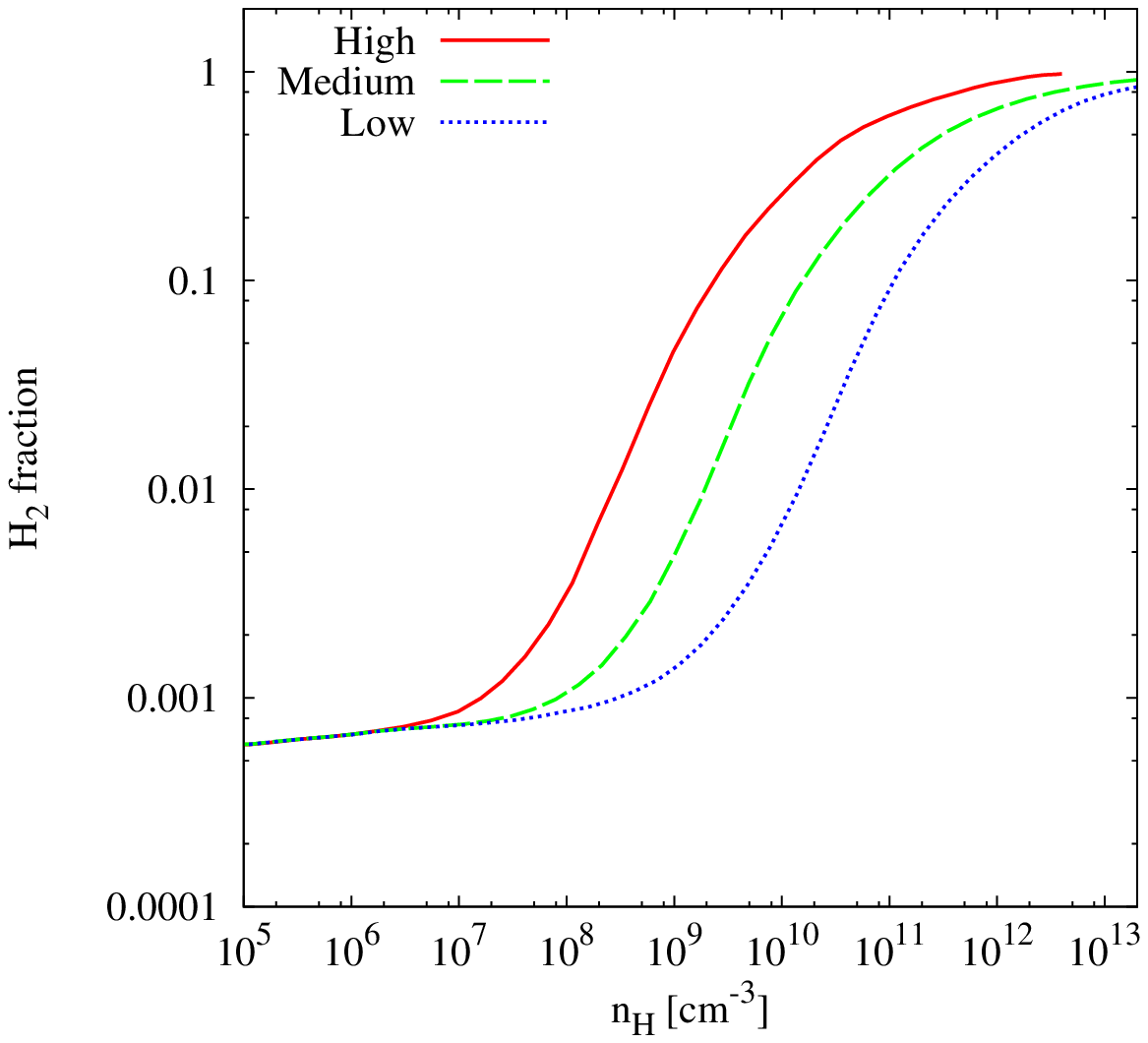} & 
\includegraphics[clip,scale=0.6]{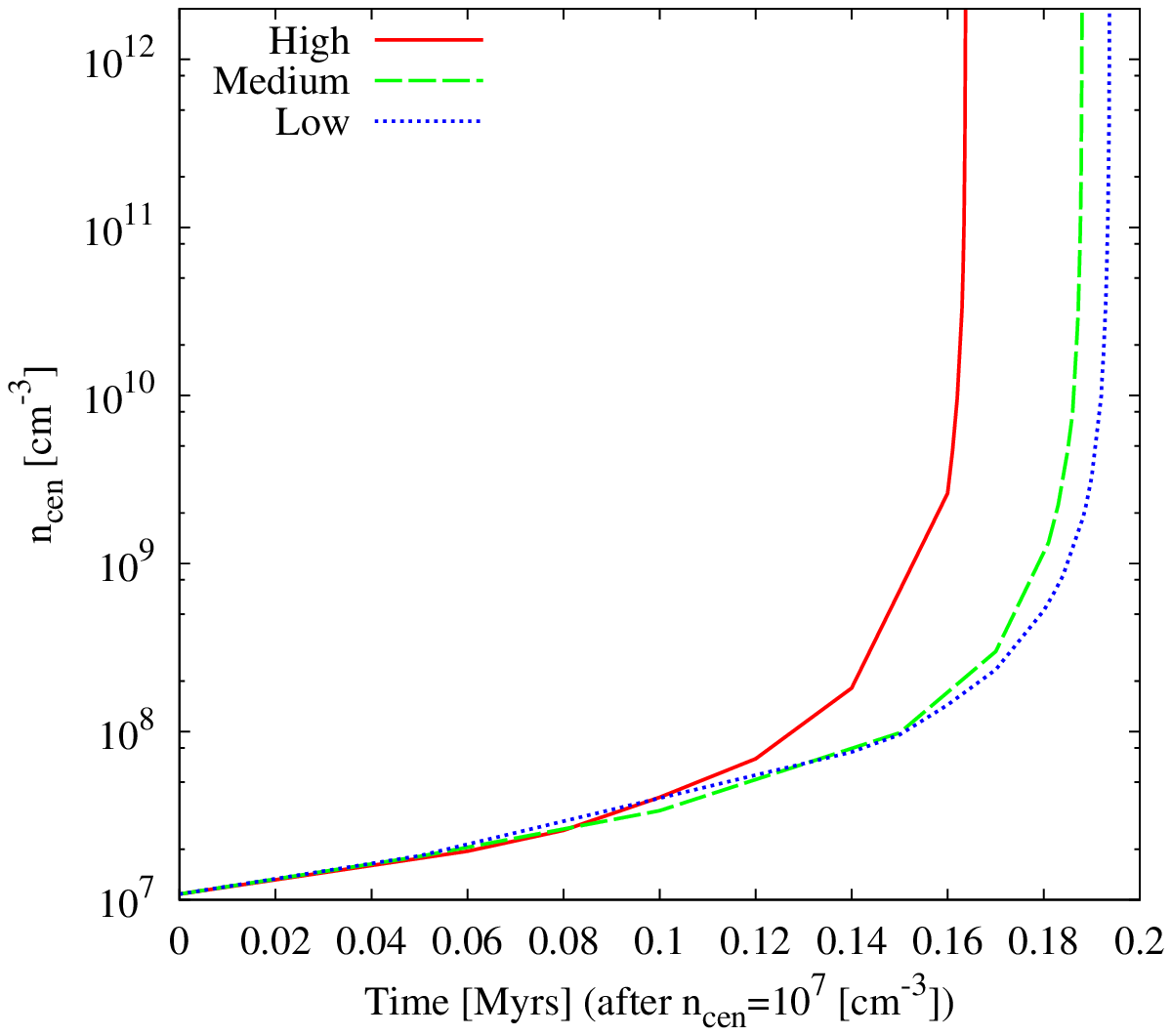}
\end{tabular}
\caption{
Left panel: Molecular fraction as a function of the gas density. 
Three lines show the runs with three reaction rates 
(see Figure \ref{f10} and Table \ref{t1}).
Right panel: The time evolution of the central density.
The horizontal axis is the elapsed time after
the central density reaches $\sim 10^{7} \ {\rm cm^{-3}}$.
}
\label{f11}
\end{center}
\end{figure}

Figure \ref{f11} shows 
the simulation results with the three reaction rates.
The left panel shows the 
\HH fraction as a function of gas density.
The density at which the cloud becomes fully-molecular differs 
more than a factor of 10. 
This is consistent with the conclusion of \cite{turk11}.
Because the molecular fraction largely determines 
the \HH line cooling rate, 
the resulting collapse time of the cloud differs by
$\Delta t \sim 30,000$ years,
depending on the choice of the reaction rate.
It is interesting that 
the time difference of the cloud collapse, $\Delta t$, is comparable to 
the difference caused by the choice of the opacity calculation method.
The difference in the molecular fraction 
becomes large at $n_{\rm H} > 10^{9} \ {\rm cm}^{-3}$, where the calculation of
the \HH line opacity is important (see Figure \ref{f6}).
Therefore, we argue that using an accurate radiative transfer method is 
as important as using an accurate reaction rate.

\section{Discussion and Conclusion}

Radiative cooling by hydrogen molecules governs
the thermal evolution of a primordial star-forming gas cloud.
At high densities, the cloud core becomes optically thick
to both the rot-vibrational lines and collision-induced continuum. 
It is necessary to estimate
the gas opacities in multiple directions
in order to calculate the radiative cooling rates accurately.
We have explicitly compared the results from 
several sets of simulations with different manners 
to calculate the optically-thick cooling rates.

\begin{figure}
\begin{center}
\begin{tabular}{cc}
\includegraphics[clip,scale=0.6]{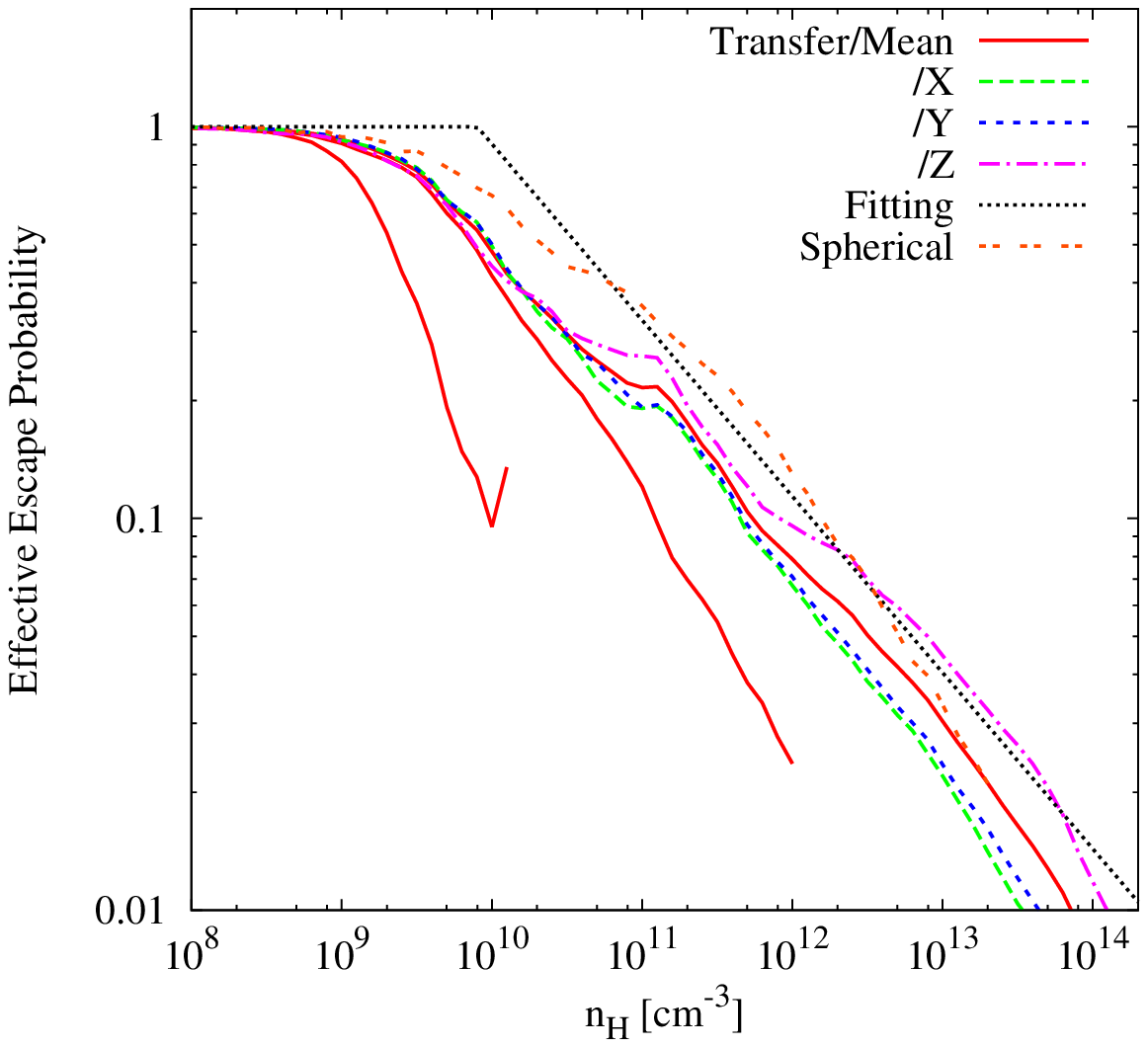} & 
\includegraphics[clip,scale=0.6]{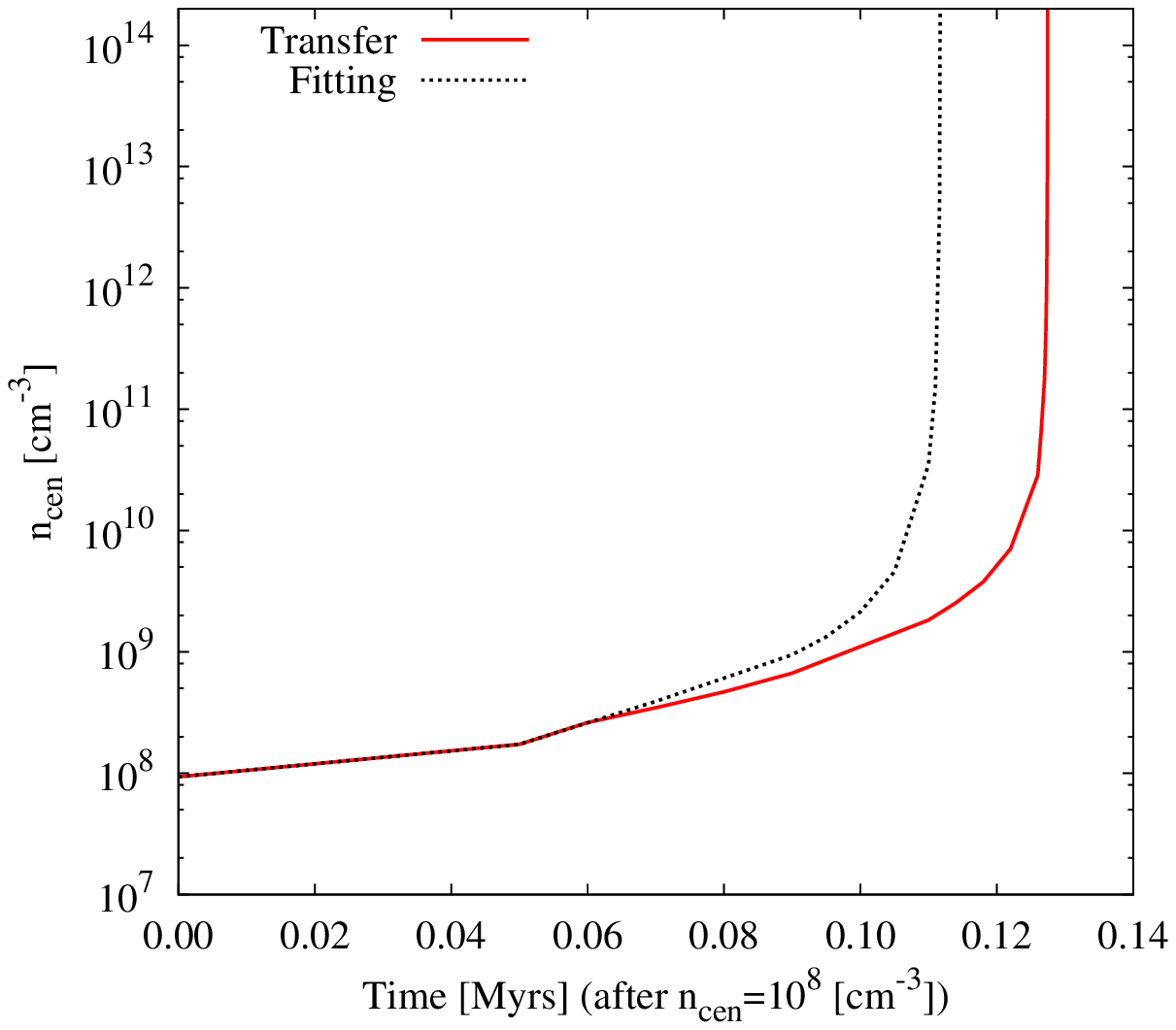}
\end{tabular}
\caption{
Left panel: The effective escape probability,
$\Lambda_{\rm thick}/\Lambda_{\rm thin}$, for ${\rm H_2}$ line
cooling for Run A. The long-dashed, short-dashed, and dot-dashed 
lines show the three orthogonal components ($x$, $y$, and $z$) 
when the central density is $n_{\rm cen} = 10^{14} \ {\rm cm^{-3}}$. 
The solid lines show the evolution of the direction-averaged mean value 
at $n_{\rm cen} = 10^{10},\ 10^{12}$, and $10^{14} \ {\rm cm^{-3}}$. 
The dotted line shows the fitting function 
given by equation (\ref{eq:line_cooling_rate_thick}).
The double-dotted line indicates the result of 
a spherical collapse calculation with 3D treatments.
Right panel: The time evolution of the central density.
The horizontal axis is the elapsed time after
the central density reaches $\sim 10^8 \ {\rm cm^{-3}}$.
}
\label{f4}
\end{center}
\end{figure}

\begin{figure}
\begin{center}
\begin{tabular}{cc}
\includegraphics[clip,scale=0.6]{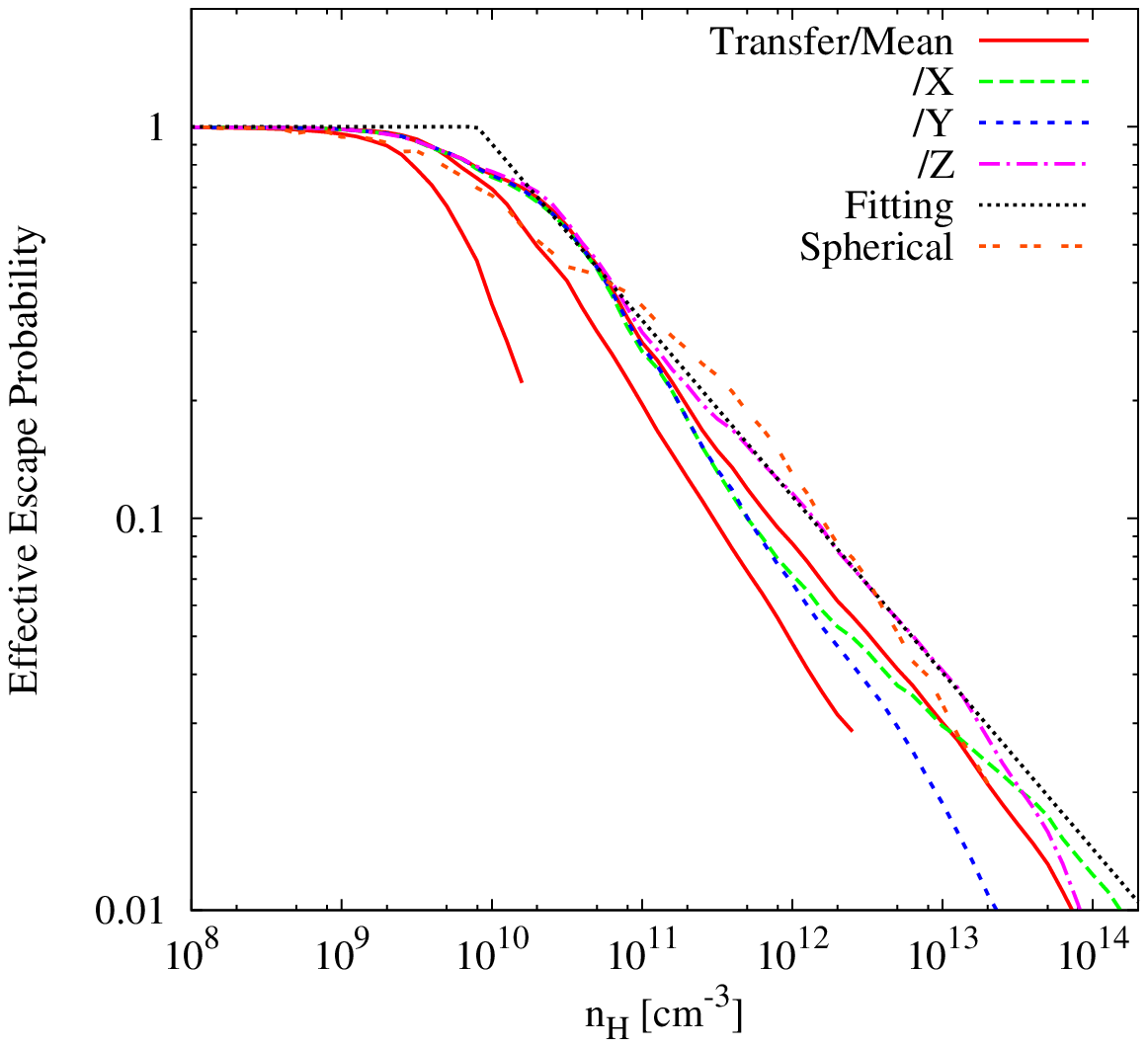} & 
\includegraphics[clip,scale=0.6]{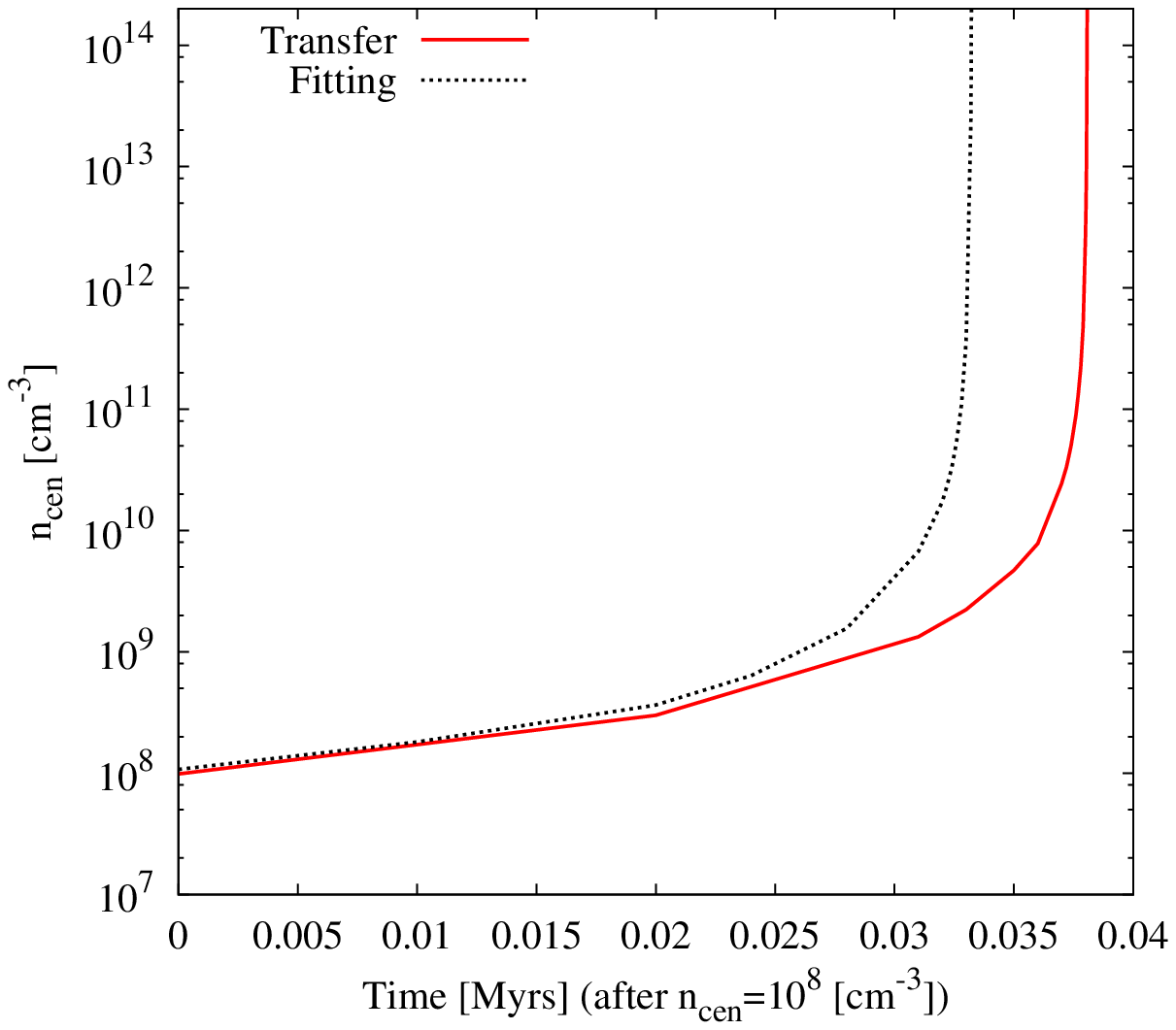}
\end{tabular}
\caption{
As for Figure \ref{f4} but for Run B.
}
\label{f6}
\end{center}
\end{figure}

\begin{figure}
\begin{center}
\begin{tabular}{cc}
\includegraphics[clip,scale=0.6]{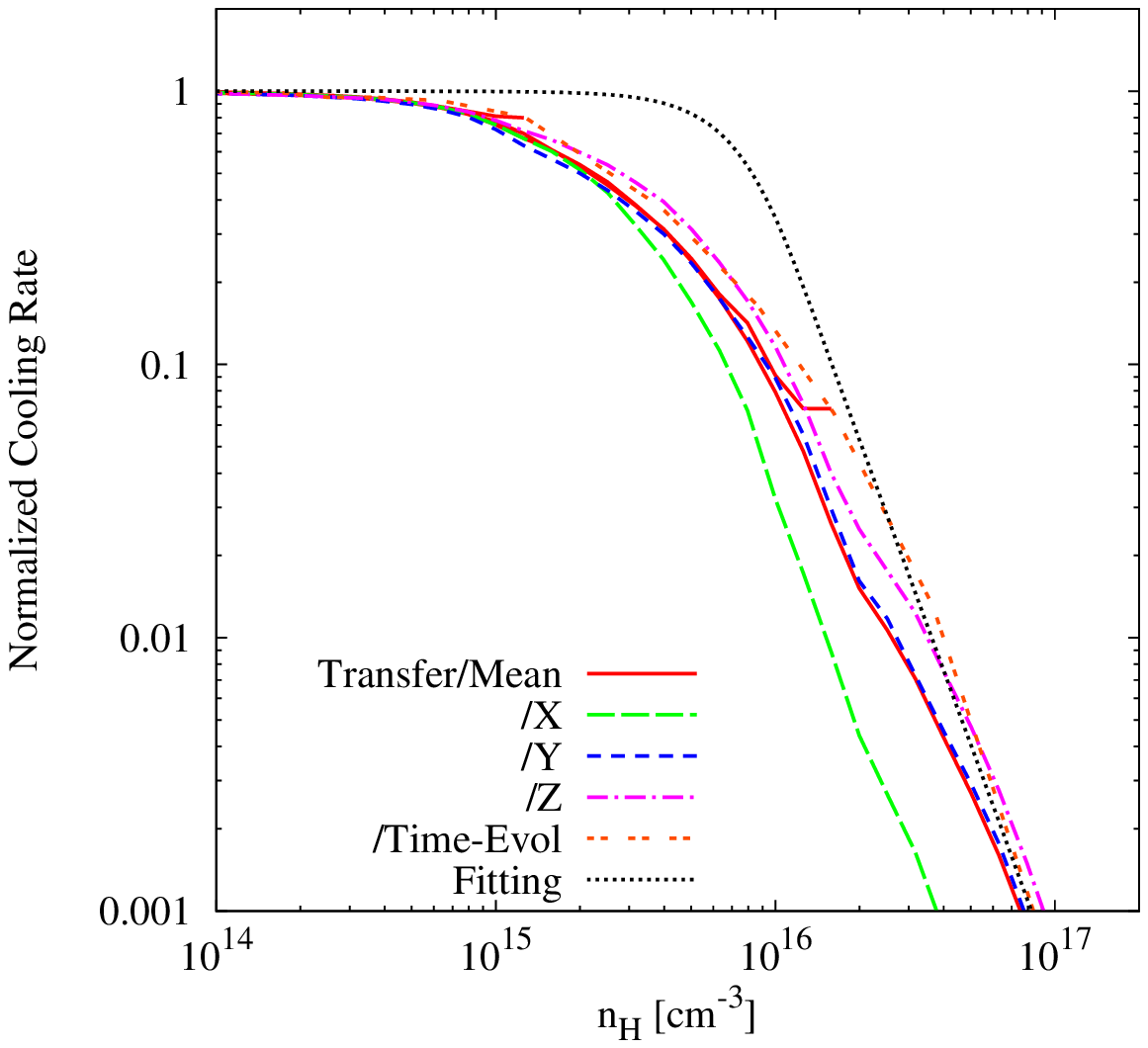} & 
\includegraphics[clip,scale=0.6]{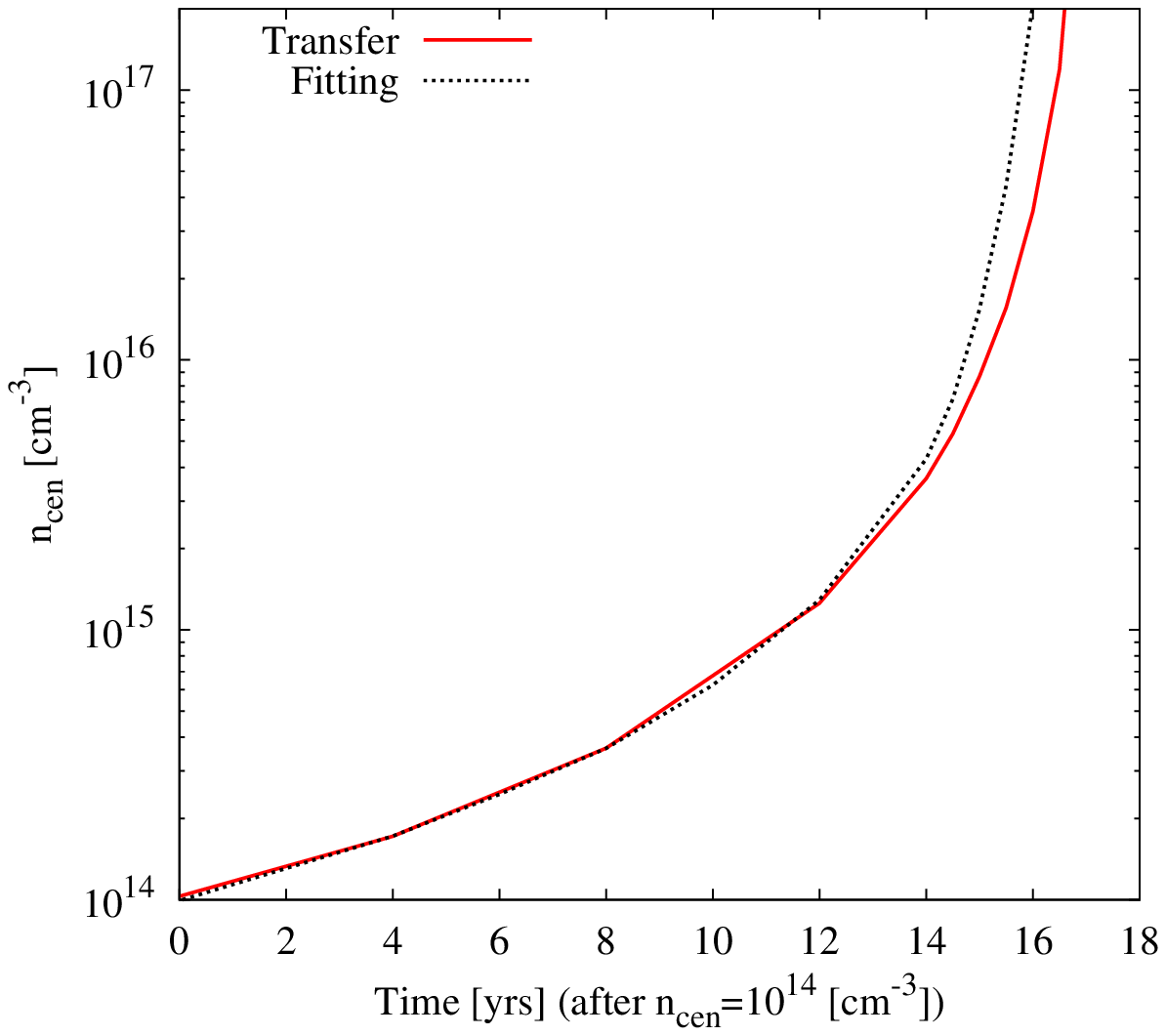}
\end{tabular}
\caption{
Left panel: The net cooling rate by ${\rm H_2}$ CIE emission for Run B. 
The long-dashed, short-dashed, and dot-dashed lines show 
the three orthogonal components ($x$, $y$, and $z$) 
when the central density is $n_{\rm cen} = 10^{17} \ {\rm cm^{-3}}$.
The solid lines show the mean values of them 
at $n_{\rm cen} = 10^{15},\ 10^{16}$, and $10^{17} \ {\rm cm^{-3}}$. 
The double-dotted line is the time-evolution of the central part. 
The dotted line shows the fitting function 
given by equation (\ref{eq:cie_cooling_rate_thick}).
Right panel: The time evolution of the central density.
The horizontal axis is the elapsed time after
the central density reaches $\sim 10^{14} \ {\rm cm^{-3}}$.
}
\label{f8}
\end{center}
\end{figure}

When non-spherical structures develop in the central region of the cloud,
photons do not escape isotropically from the dense part.
Our simulations show that the gas cloud spins up as it contracts,
and forms a flattened disk at the center. 
Then the photon escape probability not only varies
with time but is also direction-dependent.
Utilizing an ``isotropic'' fitting formula 
that is derived from spherically symmetric calculations 
over-estimates the net cooling rate and
causes the cloud core to collapse fast 
(see Figures \ref{f4}, \ref{f6}, and \ref{f8}). 
With our 3D opacity calculation, 
the photon escape fraction from the cloud core
is always smaller than given by the fitting formula.
The resulting cooling rate differs by a factor of a few to 10, 
depending on the exact density, temperature, and velocity structure. 
There is also a directional effect of the radiative transfer.
In perpendicular directions to the faces of the disk-like cloud, 
photons can easily escape. 

\begin{figure}
\begin{center}
\includegraphics[clip,scale=1]{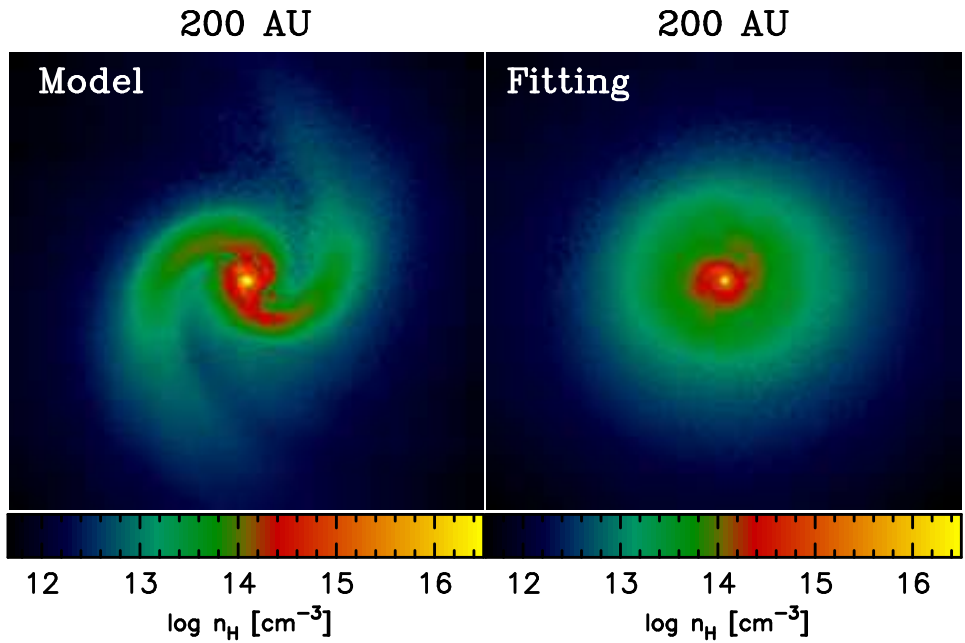}
\caption{
Projected density distributions for 
the same realization but with different opacity calculations 
at 10 years after the central density reaches 
$10^{18} \ {\rm cm^{-3}}$. 
The plotted region is $200 \ {\rm AU}$ on a side.
The left panel is for the result with our 3D opacity calculation, 
whereas the right panel is for the run with the fitting opacity for line emission.}
\label{f12}
\end{center}
\end{figure}

Details of the implementation of the optically-thick radiative cooling
affects the thermal and dynamical evolution of the cloud. 
Figure \ref{f12} shows the density distribution of 
two characteristic runs. 
We use the snapshots at 10 years after 
the central density reaches $n_{\rm cen} = 10^{18} \ {\rm cm^{-3}}$.
The left panel shows the case with our 3D radiative transfer,
whereas the right panel is for the case with the fitting formulae
\footnote{These calculations are performed with a low resolution
such that the central part does not collapse to 
much greater than $n_{\rm H} = 10^{18} \ {\rm cm^{-3}}$.
We keep the low mass resolution 
deliberately in order to follow the disk evolution over 10 years.}. 
The latter case appears rounder and 
more concentrated than former, which has a spiral structure.
Clearly, further detailed studies 
on radiative transfer effects are needed, particularly on the long-term evolution 
of the proto-stellar disk. The exact structure of the proto-stellar disk likely 
affects the disk evolution, mass accretion rate 
and fragmentation of the cloud \citep{greif12}.

A similar comparative study of radiative cooling
implementations has been done by \cite{wilkins12}
in the context of the present-day star formation.
Interestingly, an opposite trend is found 
in the case of ``polytropic'' cooling examined by \cite{wilkins12}, 
which is based on locally estimated opacities.
They show that the polytropic cooling performs 
well only in spherically symmetric cases. 
The polytropic method over-estimates the column density, 
and hence underestimates the radiative cooling rate in non-spherical cases. 
In order to calculate radiative cooling rates in 3D simulations, 
it is important to take the direction-dependence of the photon diffusion into account, 
similarly to what we conclude in the present paper.

It is clearly advantageous to use a computational method 
that is fast and robust. 
Our method of the \HH line transfer utilizes local velocity gradients 
that come with essentially no additional cost, 
because the velocity gradients are already computed 
and used in the other parts in our smoothed-particle hydrodynamics code.
For continuum photons, we need to compute the column density 
along six (or more) directions using a costly projection method devised by \cite{yoshida07}. 
In fact, the continuum opacity calculation is 
one of the most time consuming part 
in our 3D simulations.  
Nevertheless, we argue that 
it is necessary to properly take the direction-dependence into account 
in order to calculate the optically-thick radiative cooling rate accurately.
We also note that our method is based on 
the so-called escape probability method, which itself is an approximation. 
Essentially, we assume only the densest part emits continuum photons. 
It is desirable to implement fully three-dimensional radiative transfer, 
by employing advanced methods such as flux-limited diffusion or M1-closure 
(e.g., \cite{whitehouse04, levemore84})
to follow the long-term evolution of a primordial proto-stellar system.

Hydrodynamical simulations with radiative cooling 
are commonly used for 
the study of the primordial star formation.
In such simulations, 
it is important to use accurate methods 
to calculate radiative cooling rates.
For example, whether or not a proto-stellar disk fragments is determined
by the thermal and gravitational instability of the circumstellar gas.
The disk fragmentation is an important issue 
which is thought to determine 
the multiplicity and possibly the characteristic mass 
of primordial stars.
Our study clarifies the importance of multi-dimensional radiative processes 
in a primordial star-forming cloud.

\acknowledgments 

We thank Hideyuki Umeda for stimulating discussions 
and Ikko Shimizu for the technical support.
The work is supported in part by 
the Grants-in-Aid for Young Scientists (20674003:NY)  by JSPS.
The numerical calculations were in part carried out on 
SR16000 at YITP in Kyoto University and 
T2K-Tsukuba System in Center for Computational Sciences, University of Tsukuba.

\clearpage

\begin{figure}
\begin{center}
\resizebox{16cm}{!}{\includegraphics{f2.eps}}
\caption{
As for Figure \ref{f1}, but for Run B.}
\label{f2}
\end{center}
\end{figure}

\clearpage

\begin{deluxetable}{cllc}
\tablewidth{0pt}
\tablenum{1}
\tablecaption{Three-body \HH formation rate}
\tablehead{\colhead{Model} & \colhead{${\rm Rate \ Coefficient \ (cm^6 \ s^{-1})}$} & \colhead{} & \colhead{Reference}} 

\startdata
High      & $1.44 \times 10^{-26} \ T^{-1.54}$ &                     & 1 \\
Medium & $5.5  \times 10^{-29} \ T^{-1.0}$  &                     & 2 \\
Low      & $1.14 \times 10^{-31} \ T^{-0.38}$ & ($T$ $\le$ 300 {\rm K}) & 3 \\
            & $3.9  \times 10^{-30} \ T^{-1.0}$  & ($T$ $>$ 300 {\rm K})  & 3 \\
\enddata

\tablecomments{$T$ is the gas temperature in Kelvin.}
\tablerefs{(1) \cite{flower07}; (2) \cite{palla83}; (3) \cite{abel02}.}
\label{t1}
\end{deluxetable}


\begin{thebibliography}{28}
\bibitem[Abel et al.(2002)]{abel02} Abel, T., Bryan, G. L. \& Norman, M. L.  2002, Science, 295, 93
\bibitem[Borysow(2002)]{borysow02} Borysow, A.  2002, \aap, 390, 779
\bibitem[Borysow et al.(2001)]{borysow01} Borysow, A., Jorgensen, U. G. \& Fu, Y.  2001, \jqsrt, 68, 235
\bibitem[Bromm et al.(2009)]{bromm09} Bromm, V., Yoshida, N., Hernquist, L. \& McKee, C. F.  2009, \nat, 459, 49
\bibitem[Bromm \& Yoshida(2011)]{bromm11} Bromm, V. \& Yoshida, N.  2011, \araa, 49, 373
\bibitem[Clark et al.(2011)]{clark11} Clark, P. C., Glover, S. C. O., Smith, R. J., Greif, T. H., Klessen, R. S. \& Bromm, V.  2011, Science, 331, 1040
\bibitem[Flower \& Harris(2007)]{flower07} Flower, D. R. \& Harris, G. J.  2007, \mnras, 377, 705
\bibitem[Glover(2008)]{glover08} Glover, S. C. O.  2008, in AIP Conf. Proc, 990, First Star III, ed. O'Shea, B., Heger, A. \& Abel, T. (Melville, NY: AIP), 25
\bibitem[Greif et al.(2012)]{greif12} Greif, T. H., Bromm, V., Clark, P. C., Glover, S. C. O., Smith, R. J., Klessen, R. S., Yoshida, N. \& Springel, V.  2012, \mnras, 424, 399
\bibitem[Greif et al.(2011)]{greif11} Greif, T. H., Springel, V., White, S. D. M., Glover, S. C. O., Clark, P. C., Smith, R. J., Klessen, R. S. \& Bromm, V.  2011, \apj, 737, 75
\bibitem[Hosokawa et al.(2011)]{hosokawa11} Hosokawa, T., Omukai, K., Yoshida, N. \& Yorke, H. W.  2011, Science, 334, 1250
\bibitem[Jorgensen et al.(2000)]{jorgensen00} Jorgensen, U. G., Hammer, D., Borysow, A. \& Falkesgaard, J.  2000, \aap, 361, 283
\bibitem[Kitsionas \& Whitworth(2002)]{kitsionas02} Kitsionas, S. \& Whitworth, A. P.  2002, \mnras, 330, 129
\bibitem[Larson et al.(2011)]{larson11} Larson, D. et al.  2011, \apjs, 192, 16
\bibitem[Lenzuni et al.(1991)]{lenzuni91} Lenzuni, P., Chernoff, D. F. \& Salpeter, E. E.  1991, \apjs, 76, 759
\bibitem[Levemore(1984)]{levemore84} Levermore, C. D.  1984, \jqsrt, 31, 149
\bibitem[McKee \& Tan(2008)]{mckee08} McKee, C. F. \& Tan, J. C.  2008, \apj, 681, 771
\bibitem[Omukai \& Nishi(1998)]{omukai98} Omukai, K. \& Nishi, R.  1998, \apj, 508, 141
\bibitem[Palla et al.(1983)]{palla83} Palla, F., Salpeter, E. E. \& Stahler, S. W. 1983, \apj, 271, 632
\bibitem[Ripamonti et al.(2002)]{ripamonti02} Ripamonti, E., Haardt, F., Ferrara, A. \& Colpi, M. 2002, \mnras, 334, 401
\bibitem[Ripamonti \& Abel(2004)]{ripamonti04} Ripamonti, E. \& Abel, T.  2004, \mnras, 348, 1019
\bibitem[Sabano \& Yoshii(1977)]{sabano77} Sabano, Y. \& Yoshii, Y.  1977, \pasj, 29, 207
\bibitem[Silk(1983)]{silk83} Silk, J.  1983, \mnras, 205, 705
\bibitem[Springel(2005)]{springel05} Springel, V.  2005, \mnras, 364, 1105
\bibitem[Stacy et al.(2012)]{stacy12} Stacy, A., Greif, T. H. \& Bromm, V.,  2012, \mnras, 422, 290
\bibitem[Turk et al.(2009)]{turk09} Turk, M. J., Abel, T. \& O'Shea, B.  2009, Science, 325, 601
\bibitem[Turk et al.(2011)]{turk11} Turk, M. J., Clark, P., Glover, S. C. O., Greif, T. H., Abel. T., Klessen, R. \& Bromm, V.  2011, \apj, 726, 55
\bibitem[Yoshida et al.(2006)]{yoshida06} Yoshida, N., Omukai, K., Hernquist, L. \& Abel, T.  2006, \apj, 652, 6
\bibitem[Yoshida et al.(2007)]{yoshida07} Yoshida, N., Oh, S. P., Kitayama, T. \& Hernquist, L.  2007, \apj, 663, 687
\bibitem[Yoshida et al.(2008)]{yoshida08} Yoshida, N., Omukai, K. \& Hernquist, L.  2008, Science, 321, 669
\bibitem[Whitehouse \& Bate(2004)]{whitehouse04} Whitehouse, S. C. \& Bate, M. R.  2004, \mnras, 353, 1078
\bibitem[Wilkins \& Clarke(2012)]{wilkins12} Wilkins, D. R. \& Clarke, C. J.  2012, \mnras, 419, 3368
\end{thebibliography}
\end{document}